\documentclass[12pt]{article}
\pdfoutput=1
\usepackage{graphicx}
\usepackage{cite}
\usepackage{amsmath,amssymb}
\usepackage{bm}
\usepackage{subcaption}
\usepackage{caption}
\usepackage{xcolor,soul,tikz}
\usepackage{hyperref}

\title{Diffusion wall time in toroidally segmented shell \\ aka Armadillo }

\author{
	D. Abate\thanks{Corresponding author: domenico.abate@igi.cnr.it} \\
	Consorzio RFX (CNR, ENEA, INFN, Università di Padova,\\ Acciaierie Venete SpA),\\
	C.so Stati Uniti 4, 35127 Padova, Italy
	\and
	A. Corbioli \\
	Consorzio RFX (CNR, ENEA, INFN, Università di Padova,\\ Acciaierie Venete SpA),\\
	C.so Stati Uniti 4, 35127 Padova, Italy
}
\date{ April 2026}

\begin{document}
\maketitle

\begin{center}
	\textit{Accepted manuscript for Plasma Physics and Controlled Fusion (PPCF).}\\
	DOI: \href{https://doi.org/10.1088/1361-6587/ae63f7}{10.1088/1361-6587/ae63f7}
\end{center}

\begin{abstract}
An analytical expression for the diffusion wall time of a toroidally 
segmented conducting shell (the Armadillo configuration) is derived by 
extending the continuous-shell formulation to include the non-axisymmetric 
current pattern imposed by the presence of toroidal gaps. The segmentation 
constrains the toroidal current to follow a standing-wave structure that 
vanishes at the gap locations, introducing a correction to the effective 
resistivity that grows quadratically with the number of gaps and competes 
with the intrinsic toroidal scale of the mode. As a result, the wall time 
decreases rapidly for low toroidal-number modes, more gradually for 
intermediate ones, and only for sufficiently large segmentation in the 
high-$n$ regime. The analytical formula shows agreement within $10\%$ 
against 3D electromagnetic numerical calculations. The resulting expression 
provides a compact tool for estimating the wall time of segmented conducting 
structures surrounding the plasma, with direct applications to MHD stability 
and control in both RFPs and tokamaks.
\end{abstract}


\section{Introduction}
In magnetic confinement fusion devices, the diffusion of magnetic fields through
the conducting structures surrounding the plasma plays a central role in magneto‑hydrodynamic (MHD) stability and control. In particular, the so-called wall time $\tau_w$ sets the characteristic timescale for magnetic field diffusion through the wall and strongly influences the growth rate of
resistive wall modes (RWMs)\cite{strait2003resistive, chu2010stabilization,
pustovitov2015plasma}. The commonly used expression for the wall time:
\begin{equation}
	\tau_w \equiv \mu_0 r_w t_w \eta_w^{-1}
	\label{eq:wall_time}
\end{equation}
refers to the magnetic field diffusion time scale through a thin continuous cylindrical resistive shell of radius $r_w$, thickness $t_w$ and with constant electrical resistivity $\eta_w$. The wall time is also equivalently interpreted as the characteristic decay time of eddy‑current patterns induced by a RWM \cite{reimerdes2006cross}. Regardless of the adopted physical picture, this widely adopted formula was firstly introduced, to the best of our knowledge, by C. G. Gimblett in the Appendix of \cite{gimblett1986free} who obtained the decay rate by matching vacuum magnetic fields across a resistive thin wall. This formulation naturally introduces the possibility of different toroidal and poloidal resistivities and leads to the so‑called "Gimblett effective resistivity", which mixes different poloidal and toroidal wall resistivities (for example due to a bellows construction)\cite{gimblett1986free}. 

A generalization of this approach to perturbations of arbitrary helicity $(m,n)$ - where $m$ is the poloidal and $n$ the toroidal wavenumbers- was presented in \cite{zancaFC57}. There, the diffusion time $\gamma(m,n)^{-1}$ is obtained from the jump of the radial derivative of the magnetic flux across the wall. In the limit $n \to 0$, the resulting expression reduces to the classical vertical field penetration time. Importantly, the commonly used wall time $\tau_w$ - or "long" time constant as named in \cite{gimblett1986free} -  actually corresponds to twice of the vertical field penetration time. This relation is well known for circular shells \cite{isernia2019cross} and follows from the classical cylindrical–toroidal correspondence. However, these characteristic times are sensitive to the specific vessel geometry and elongation: in particular, the derivation of eddy‑current eigenmodes and decay times for elliptical cross-section walls \cite{chukashev2025toroidal} further contributes to the identification of these characteristic scales.

The distinction between the various characteristic wall times has been
examined in previous analytical studies. In particular, in \cite{dialetis1991diffusion} is shown that the magnetic field penetration through a toroidal conducting shell involves multiple time scales, only one of which corresponds to the physical diffusion time. More recent analyses \cite{pustovitov2008general, pustovitov2018reaction} further clarified that the wall time $\tau_w$ commonly used in RWM theory is primarily a convenient normalization parameter, while the actual decay time of a perturbation with poloidal number $m\geq 1$ scales as $\tau_w/(2m)$ for axisymmetric perturbations 
and a circular shell \cite{pustovitov2008general, 
pustovitov2018reaction, chukashev2025toroidal}.  A similar distinction between the characteristic wall time and the geometric multipliers entering the RWM dispersion relation is discussed in \cite{pustovitov2023general}, where the normalization differs from the conventional $\tau_w$ (see Eqs. (57) and (71) therein). These distinctions are often irrelevant when $\tau_w$ is used as a normalization parameter for example in RWM stability calculations where the dispersion relation naturally involves the dimensionless combination $\gamma \tau_w$. Nevertheless, the absolute value of $\tau_w$ becomes relevant when growth rates are expressed in physical units for designing feedback control systems in both Tokamak and reversed field pinch (RFP) devices. 

In RFP devices, a toroidal passive stabilizing shell (PSS) is employed to slow down the broad spectrum of MHD instabilities to resistive time scales. In practice, the PSS includes toroidal and poloidal gaps that interrupt the net circulation of eddy currents. A new configuration of interest is the toroidally segmented shell or Armadillo configuration (inspired by the segmented structure of the animal’s shell). This concept emerged during the design of the RFX‑mod2 experiment \cite{marrelli2019optimization}, where the PSS is placed directly in vacuum inside the vacuum‑tight support structure (VTSS) \cite{peruzzo2023new}. In this environment, electrical insulation of the PSS becomes critical and difficult to guarantee because of the presence of a diffused plasma even in the "ideally" vacuum regions\cite{cordaro2024electrical}. Introducing multiple toroidal gaps reduces the induced electromotive force on each segment, easing insulation requirements and simplifying mechanical assembly.

In the context of vertical plasma position control, the impact of toroidal gaps on the  eddy-current distribution and the associated decay time in a resistive shell was first modelled in \cite{nagayama1984feedback} where an analytical 
estimate of the gap-induced correction to the shell resistance was derived for the axisymmetric ($m=1$, $n=0$) mode.
In addition, segmentation reduces the magnitude of the PSS error fields, which is beneficial for MHD control.

The aim of this work is to derive an analytical expression for the wall time of
a toroidally segmented thin shell. The key step is to represent the
segmentation‑induced modulation of the toroidal surface current as a standing
wave constrained by the gap positions, which introduces a segmentation
wavenumber proportional to the number of gaps. This additional spectral
component enhances the poloidal dissipation channel and modifies the effective
resistivity by adding a segmentation term to the classical Gimblett expression.
The resulting wall time retains the structure of the continuous‑shell formula
but incorporates segmentation through the modified resistivity (Section
\ref{sec:math}). The analytical model highlights how segmentation competes with the intrinsic $(m,n)$ structure of each harmonic (Section \ref{sec:results}), and its predictions are consistent with 3D electromagnetic numerical calculations. The
resulting formulation provides a compact analytical tool that can be applied to
both RFPs and tokamaks, where discrete conducting structures (e.g. shell,
blanket modules) can affect MHD stability and control (Section
\ref{sec:conclusions}).


\section{Mathematical model}
\label{sec:math}
The problem is solved in vacuum, where the magnetic field of given helicity
$(m,n)$ can be produced either inside the shell (e.g. by plasma current) or outside it (e.g. by coils or other conductors). Throughout this work, the presence of plasma is neglected: this would 
modify the boundary conditions at the plasma surface, introducing 
a coupled plasma-wall problem which is beyond the scope of the 
present analysis. We consider a thin
cylindrical conducting shell of radius $r_w$ and thickness $t_w \ll r_w$,
characterized by an electrical resistivity with components $\eta_\phi$
and $\eta_\theta$ in the toroidal and poloidal directions, respectively. The
system is described in cylindrical coordinates $(r,\theta,\phi)$, where
$\phi = z/R_0$ mimics the toroidal angle. In the following, the magnetic field is decomposed as 
\(\mathbf{B} = \mathbf{B}_0 + \mathbf{b}\), where \(\mathbf{B}_0\) denotes 
the equilibrium field and \(\mathbf{b}\) the perturbation.
The magnetic field evolution is described by Maxwell equations in the
quasi-static approximation:
\begin{equation}
\nabla \times \mathbf{E} = -\frac{\partial \mathbf{B}}{\partial t}, \qquad
\nabla \times \mathbf{B} = \mu_0 \mathbf{J}, \qquad
\nabla \cdot \mathbf{B} = 0.
\end{equation}
In the vacuum regions ($r < r_w,\, r > r_w$), where $\mathbf{J} = \mathbf{0}$,
the magnetic field is both solenoidal and curl-free. The curl-free condition
$\nabla \times \mathbf{b} = \mathbf{0}$ allows to write the magnetic field as the gradient of a scalar potential:
\begin{equation}
	\mathbf{b} = \nabla\Phi,
	\label{eq:BgradPhi}
\end{equation}
and the divergence-free condition implies that $\Phi$ satisfies Laplace’s
equation:
\begin{equation}
\nabla^2 \Phi = 0.
	\label{eq:laplace}
\end{equation}

Exploiting linearity and the spatial periodicities of the system, the
scalar potential is written as a sum of harmonics:
\begin{equation}
\Phi(r,\theta,\phi,t) = \tilde{\Phi}(r)\,e^{i(m\theta - n\phi)}\,e^{-\gamma t}.
\label{eq:phi}
\end{equation}
Substituting into \eqref{eq:laplace} yields the modified Bessel equation for the
radial function $\tilde{\Phi}(r)$:
\begin{equation}
	\frac{\mathrm{d}^2\tilde{\Phi}}{\mathrm{d}r^2}
	+ \frac{1}{r}\frac{\mathrm{d}\tilde{\Phi}}{\mathrm{d}r}
	- \left(\frac{m^2}{r^2} + \frac{n^2}{R_0^2}\right)\tilde{\Phi} = 0.
\end{equation}

The regular solution in the inner vacuum region ($0 < r < r_w$) is
$\tilde{\Phi}_{\rm in}(r) = C\,I_m(kr)$, whereas the decaying solution in the
outer vacuum region ($r > r_w$) is $\tilde{\Phi}_{\rm out}(r) = D\,K_m(kr)$,
where $I_m$ and $K_m$ are the modified Bessel functions of the first and second
kind of order $m$, respectively.
Continuity of the radial magnetic field at the wall ($r = r_w$) requires
\begin{equation}
\left.\frac{\mathrm{d}\tilde{\Phi}_{\rm in}}{\mathrm{d}r}\right|_{r_w}
= \left.\frac{\mathrm{d}\tilde{\Phi}_{\rm out}}{\mathrm{d}r}\right|_{r_w}
\equiv b_{rw} = b_r(r_w),
\end{equation}
which allows to express the potential on both sides in terms of the radial
field amplitude:
\begin{equation}
\tilde{\Phi}_{\rm in}(r_w)
= b_{rw}\,\frac{r_w}{u}\,\frac{I_m(u)}{I_m'(u)},
\qquad
\tilde{\Phi}_{\rm out}(r_w)
= b_{rw}\,\frac{r_w}{u}\,\frac{K_m(u)}{K_m'(u)},
\qquad u = \frac{n r_w}{R_0}.
\end{equation}
where primes denote derivatives with respect to the argument, evaluated at
$u = n r_w / R_0$.

In the thin–wall approximation ($t_w \ll r_w$), the current density inside the
shell is assumed to have no radial component ($J_r = 0$) and to be uniform
across the thickness. The magnetic field is likewise taken to vary negligibly
through the wall, so that $B_r$ is evaluated at $r = r_w$. Under these
assumptions, the current in the wall is represented as a surface current $\mathbf{K} = \int_0^{t_w} \mathbf{J}\,\mathrm{d}r = (K_\theta, K_\phi)$ following the classical thin-shell treatment \cite{morozov1966}. The discontinuity of the tangential magnetic field across the wall is then
related to the surface current through the standard jump condition
\begin{equation}
\hat{r} \times (\mathbf{B}^{\rm out} - \mathbf{B}^{\rm in}) = \mu_0\mathbf{K}
\end{equation}
which gives
\begin{equation}
\mu_0 K_\phi = \Delta B_\theta, \qquad
\mu_0 K_\theta = -\Delta B_\phi.
\end{equation}
Using the field components, the surface currents are linked to the jump in the
scalar potential
$\Delta\tilde{\Phi}_w = \tilde{\Phi}_{\rm out}(r_w) - \tilde{\Phi}_{\rm in}(r_w)$:
\begin{equation}
K_\phi = \frac{i m}{\mu_0 r_w}\,\Delta\tilde{\Phi}_w, \qquad
K_\theta = \frac{i n}{\mu_0 R_0}\,\Delta\tilde{\Phi}_w.
\label{eq:K_theta_phi}
\end{equation}
The ratio of these two expressions gives
\begin{equation}
K_\theta = \frac{n r_w}{m R_0}\,K_\phi = \frac{u}{m}\,K_\phi,
\end{equation}
which satisfies the surface current continuity condition
$\nabla \cdot \mathbf{K} = 0$.

The potential jump is evaluated exactly using the Wronskian identity
$I_m(u)K_m'(u) - I_m'(u)K_m(u) = -1/u$:
\begin{equation}
\Delta\tilde{\Phi}_w
= \frac{b_{rw}\,r_w}{u}
\frac{I_m'(u)\,K_m(u) - I_m(u)\,K_m'(u)}
{I_m'(u)\,K_m'(u)}
= \frac{b_{rw}\,r_w}{u^2\,I_m'(u)\,K_m'(u)}.\end{equation}
We therefore introduce the geometric factor
\begin{equation}
G_m(u) = -\frac{1}{u^2\,I_m'(u)\,K_m'(u)},
\label{eq:Gm_exact}
\end{equation}
which is positive for all $u > 0$ since $I_m' > 0$ and $K_m' < 0$, so that
\begin{equation}
\Delta\tilde{\Phi}_w = -\,b_{rw}\,r_w\,G_m(u).
\label{eq:DeltaPhi}
\end{equation}
In the limit $u \to 0$ ($n=0$), using the small-argument asymptotics 
	$I_m'(u) \sim u^{m-1}/[2^m(m-1)!]$ and 
	$K_m'(u) \sim -m(m-1)!\,2^{m-1}/u^{m+1}$, one finds 
	$u^2 I_m'(u)K_m'(u) \to -m/2$, so that
	\begin{equation}
		G_m(0) = \frac{2}{m},
	\end{equation}
	which recovers the standard decay time $\tau_w^{\rm (cont)}(m,0) = \tau_w/(2m)$ for axisymmetric 
	perturbations of a circular shell \cite{pustovitov2008general, 
		pustovitov2018reaction, chukashev2025toroidal}, as shown explicitly 
		in the following section, see relation \eqref{eq:tauw_vertical}.
\subsection{Continuous shell}
The characteristic decay rate $\gamma$ is obtained by evaluating the radial
component of Faraday's law:
\begin{equation}
	\left.(\nabla \times \mathbf{E})_r\right|_{r=r_w}
	= -\,\left.\frac{\partial B_r}{\partial t}\right|_{r=r_w}.
\end{equation}
which in cylindrical coordinates, with the assumed modal dependence, reduces to:
\begin{equation}
\frac{im}{r_w}\,E_\phi + \frac{in}{R_0}\,E_\theta = \gamma\,b_{rw}.
\label{eq:faraday_cyl}
\end{equation}
Integrating Ohm’s law across the wall thickness yields the surface currents
\begin{equation}
	K_\phi = \int_0^{t_w} J_\phi\,\mathrm{d}r = \frac{t_w}{\eta_\phi} E_\phi,
	\qquad
	K_\theta = \int_0^{t_w} J_\theta\,\mathrm{d}r = \frac{t_w}{\eta_\theta} E_\theta.
\end{equation}
With \eqref{eq:K_theta_phi} this transforms \eqref{eq:faraday_cyl} into 
\begin{equation}
\gamma\,b_{rw}
= -\frac{m^2\eta_\phi + u^2\eta_\theta}{\mu_0\,t_w\,r_w^2}\,\Delta\tilde{\Phi}_w.
\end{equation}
By using \eqref{eq:DeltaPhi}, the radial field $b_{rw}$ cancels out yielding the decay rate:
\begin{equation}
	\gamma(m,n)
	= \frac{m^2\eta_\phi + u^2\eta_\theta}{\mu_0\,t_w\,r_w}\,G_m(u).
\label{eq:gamma_exact}
\end{equation}
The corresponding wall time for the continuous shell is
\begin{equation}
\tau_w^{\rm (cont)}(m,n)
= \gamma(m,n)^{-1} = \frac{\mu_0\,t_w\,r_w}{(m^2\eta_\phi + u^2\eta_\theta)\,G_m(u)}.
\label{eq:tau_cont}
\end{equation}
For an axisymmetric vertical field $(m=1,\,n=0)$, using $G_1(0) = 2$ and $\eta_\phi = \eta_w$, leads to:
\begin{equation}
\tau_w^{\rm (cont)}(1,0) = \frac{\mu_0\,t_w\,r_w}{2\eta_w},
\label{eq:tauw_vertical}
\end{equation}
which shows that the well-known and commonly adopted definition in \eqref{eq:wall_time} - corresponding to the "long" time constant of the wall in \cite{gimblett1986free}- is twice the characteristic diffusion time associated with the penetration of a vertical ($m=1,n=0$) magnetic perturbation.
This combination of $\eta_\phi$ and $\eta_\theta$ is equivalent to the effective wall resistivity introduced in \cite{gimblett1986free} for a given $(m,n)$, up to the factor $m^2 + u^2$.

\subsection{Segmented shell (Armadillo)}
\begin{figure}[tb!]
	\centering
	\includegraphics[width=1.\textwidth]{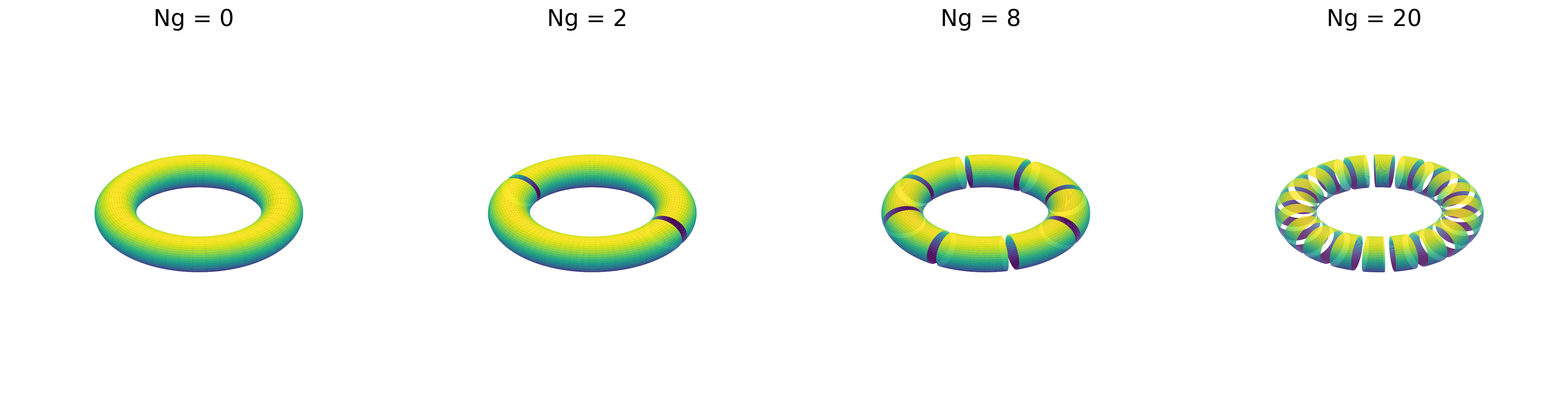}
	\caption{Continuous toroidal shell ($N_g=0$) and segmented one (Armadillo) where $N_g$ is the number of gaps.}
	\label{fig:armadillo}
\end{figure}
The following derivation applies to $m \geq 1$: $m=1$ 
	modes dominate the MHD spectrum in RFP devices, while higher-$m$ 
	harmonics ($m \geq 2$) are important for error-field correction and 
	resistive wall mode dynamics in tokamaks. For $m=0$, the jump 
	condition \eqref{eq:K_theta_phi} gives $K_\phi = 0$ identically, 
	so there is no toroidal surface current to modulate and the 
	segmentation has no effect within the present framework; that case 
	would require a separate treatment.

Consider a shell divided into $N_g$ identical toroidal segments (Fig.~\ref{fig:armadillo}). The
segmentation constrains the toroidal surface current to vanish at the gap
locations $\phi_j = 2\pi j/N_g$, which consequently modulates the $(m,n)$
Fourier component of the current.
The toroidal surface current takes the form
\begin{equation}
K_\phi(\theta,\phi,t)
= \mathcal{K}_\phi(t)\,e^{i(m\theta-n\phi)}\,w(\phi,t),
\end{equation}
where the modulation function $w(\phi,t)$ enforces $w(\phi_j,t)=0$ at the gaps.
Modeling $w$ as the fundamental standing wave compatible with the gap pattern,
\begin{equation}
w(\phi,t) = \sin(k_\phi\phi)\,T(t),
\qquad k_\phi = \frac{N_g}{2},
\qquad k_s = \frac{k_\phi}{R_0}.
\end{equation}
follows from the natural boundary conditions at the gap edges and from the dominance of the lowest eigenmode in the thin–gap limit. 

Enforcing current continuity $\nabla\cdot\mathbf{K}=0$ on the segmented shell,
the poloidal surface current becomes
\begin{equation}
K_\theta
= \frac{r_w}{mR_0}\,\mathcal{K}_\phi\,e^{i(m\theta-n\phi)}
\bigl[i n\,w(\phi,t) - \partial_\phi w(\phi,t)\bigr].
\end{equation}
Averaging the squared magnitudes over one toroidal period:
\begin{equation}
\langle|K_\phi|^2\rangle_\phi = \tfrac{1}{2}|\mathcal{K}_\phi|^2|T|^2,
\qquad
\langle|K_\theta|^2\rangle_\phi
= \left(\frac{r_w}{mR_0}\right)^2
\tfrac{1}{2}|\mathcal{K}_\phi|^2|T|^2\,(n^2 + k_\phi^2).
\end{equation}
The average Joule dissipation per unit area is 
\begin{equation}
P = \langle|K_\phi|^2\rangle_\phi
\left[\eta_\phi + \eta_\theta\left(\frac{r_w}{mR_0}\right)^2(n^2+k_\phi^2)\right].
\end{equation}
Defining the reference current amplitude consistently with the continuous case,
$K_{\rm ref}^2 = \langle|K_\phi|^2\rangle_\phi\,(m^2+u^2)/m^2$, the effective
resistivity of the segmented shell follows from $P \equiv K_{\rm ref}^2\,\eta_{\rm eff}$:
\begin{equation}
\eta_{\rm eff}(m,n;N_g)
= \frac{m^2\eta_\phi + u^2\eta_\theta}{m^2 + u^2}
+ \frac{\eta_\theta}{m^2+u^2}
\left(\frac{r_w}{R_0}\right)^2\!\left(\frac{N_g}{2}\right)^2
= \frac{m^2\eta_\phi + u^2\eta_\theta 
	+ \eta_\theta\,(r_w k_s)^2}{m^2 + u^2}.
\label{eq:eta_eff_armadillo}
\end{equation}

By analogy with \eqref{eq:tau_cont}, the characteristic wall time for the
Armadillo shell can be expressed as
\begin{equation}
\tau_w(m,n;N_g)
= \frac{\mu_0\,t_w\,r_w}{(m^2+u^2)\,G_m(u)\,\eta_{\rm eff}(m,n;N_g)},
\qquad u = \frac{nr_w}{R_0}.
\label{eq:walltime_arma}
\end{equation}
In the continuous limit ($N_g = 0$), relation \eqref{eq:walltime_arma} reduces exactly to \eqref{eq:tau_cont}.
The quadratic scaling of the 
effective resistivity with the number of gaps is consistent with the 
earlier circuit-model result of \cite{nagayama1984feedback}, derived for the 
axisymmetric $(m=1,\,n=0)$ mode; the present formulation extends that 
result to arbitrary helicity $(m,n)$ and anisotropic wall resistivity.

\section{Results}
\label{sec:results}
\begin{figure}[tb!]
    \centering
    \includegraphics[width=.475\textwidth]{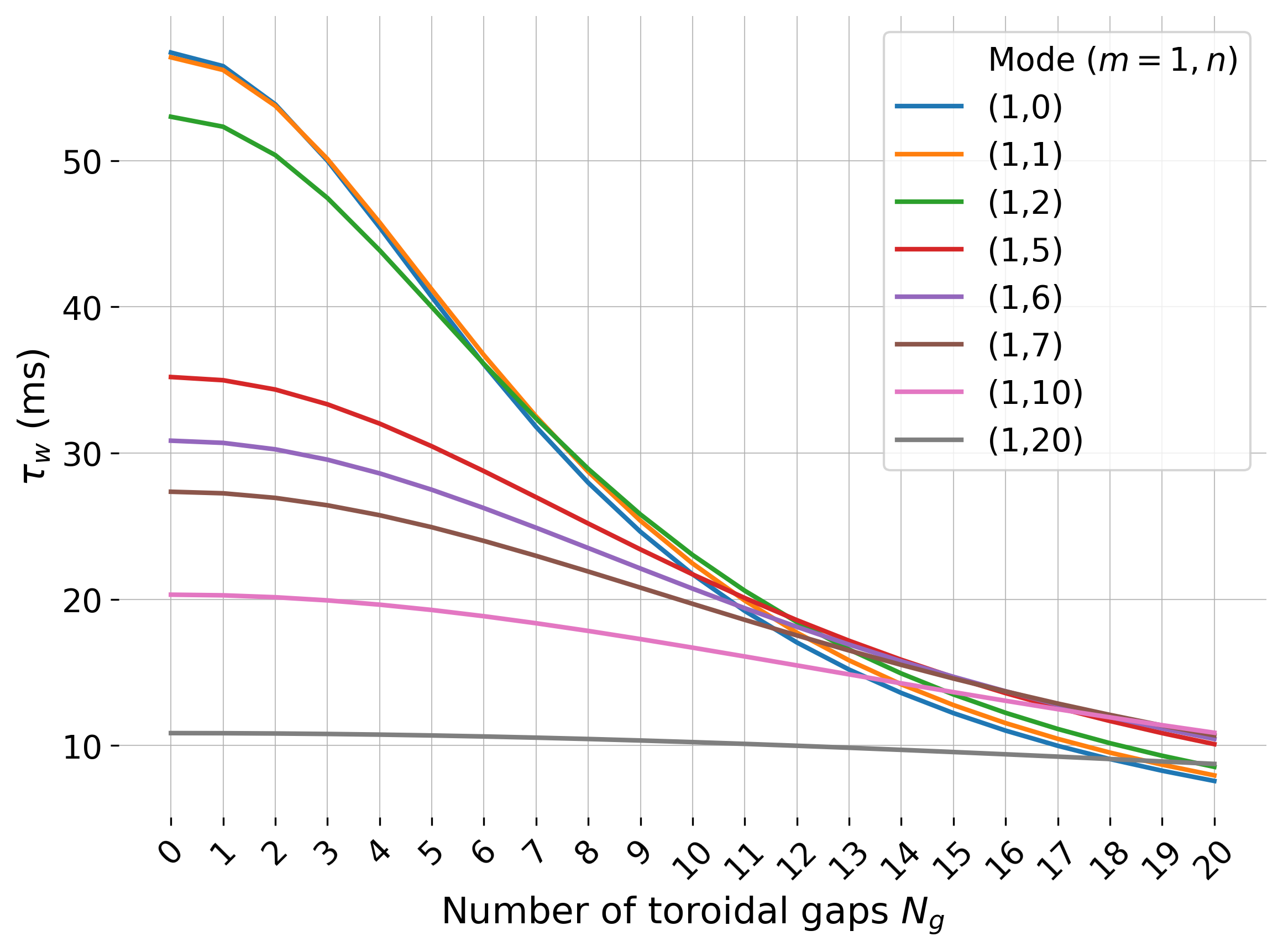}
    	\includegraphics[width=.475\textwidth]{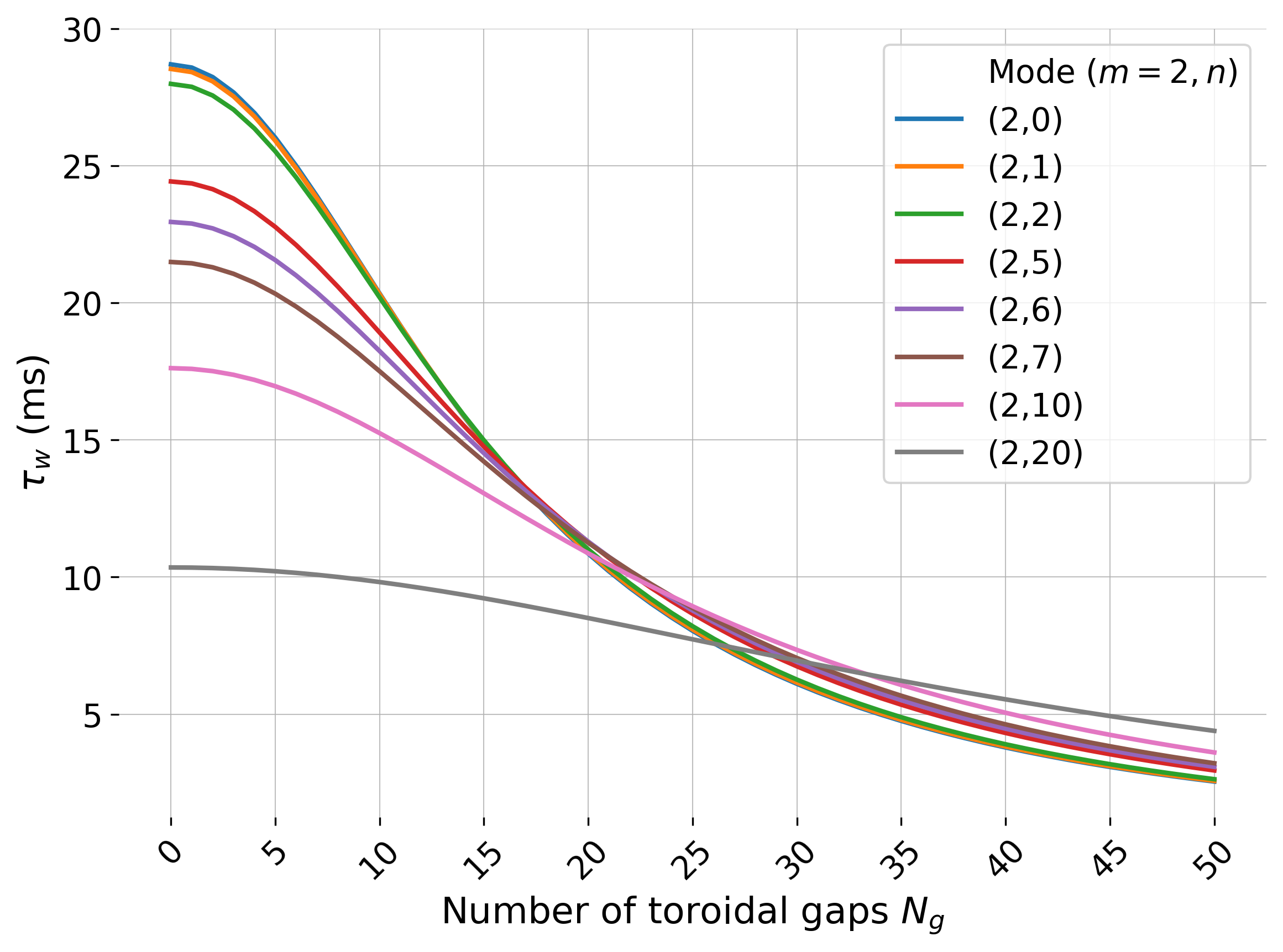}
    \caption{Wall time given by the analytical model for different $(m=1,n)$ (left) and $(m=2,n)$ (right) as a function of different number of gaps $N_g$.}
    \label{fig:walltime_ok}
\end{figure}

The parameters used throughout ($r_w = 0.5115$~m, $t_w = 3$~mm, 
$R_0 = 1.995$~m, $\eta_\phi = \eta_\theta = 1.68\times10^{-8}$~$\Omega$m, $\varepsilon_w \equiv r_w/R_0 \approx 0.26$) refer to the passive stabilizing shell of the RFX-mod2 experiment.
We focus on the $m=1$ modes, which dominate the RFP magnetic configuration and
are therefore the most relevant harmonics for assessing the impact of toroidal
segmentation in the Armadillo concept. The segmented shell wall time in  \eqref{eq:walltime_arma}, is evaluated for different $(m=1,n)$ modes and for a range of toroidal gap numbers $N_g$ (Table~\ref{tab:results_compact} and Fig.~\ref{fig:walltime_ok}). As expected from  \eqref{eq:walltime_arma}, increasing the number of gaps $N_g$ enhances the segmentation term proportional to $(N_g/2)^2$, thereby reducing the wall time relative to the continuous–shell value.
\begin{table}[tb!]
	\centering
	\caption{Wall time $\tau_w$ (ms) for $m=1$ as a function of toroidal mode number $n$ and number of toroidal gaps $N_g$.}
	\begin{tabular}{c|c|c|c|c|c|c|c|c|c}
		$n$ & $N_g=0$ & $1$ & $2$ & $4$ & $6$ & $8$ & $10$ & $15$ & $20$ \\
		\hline
		0  & 57.39 & 56.46 & 53.85 & 45.44 & 36.06 & 27.97 & 21.71 & 12.22 & 7.58 \\
		1  & 57.06 & 56.20 & 53.75 & 45.77 & 36.69 & 28.72 & 22.45 & 12.77 & 7.96 \\
		2  & 53.00 & 52.32 & 50.38 & 43.87 & 36.09 & 28.92 & 23.03 & 13.49 & 8.54 \\
		5  & 35.20 & 34.98 & 34.35 & 32.02 & 28.76 & 25.18 & 21.71 & 14.67 & 10.10 \\
		6  & 30.85 & 30.70 & 30.26 & 28.61 & 26.24 & 23.50 & 20.73 & 14.70 & 10.45 \\
		7  & 27.36 & 27.25 & 26.94 & 25.75 & 23.99 & 21.90 & 19.69 & 14.58 & 10.70 \\
		10 & 20.32 & 20.27 & 20.14 & 19.63 & 18.84 & 17.84 & 16.69 & 13.65 & 10.88 \\
		20 & 10.85 & 10.84 & 10.82 & 10.75 & 10.62 & 10.45 & 10.23 & 9.56  & 8.74 \\
	\end{tabular}
	\label{tab:results_compact}
\end{table}

\begin{figure}[tb!]
    \centering
    \includegraphics[width=.6\textwidth]{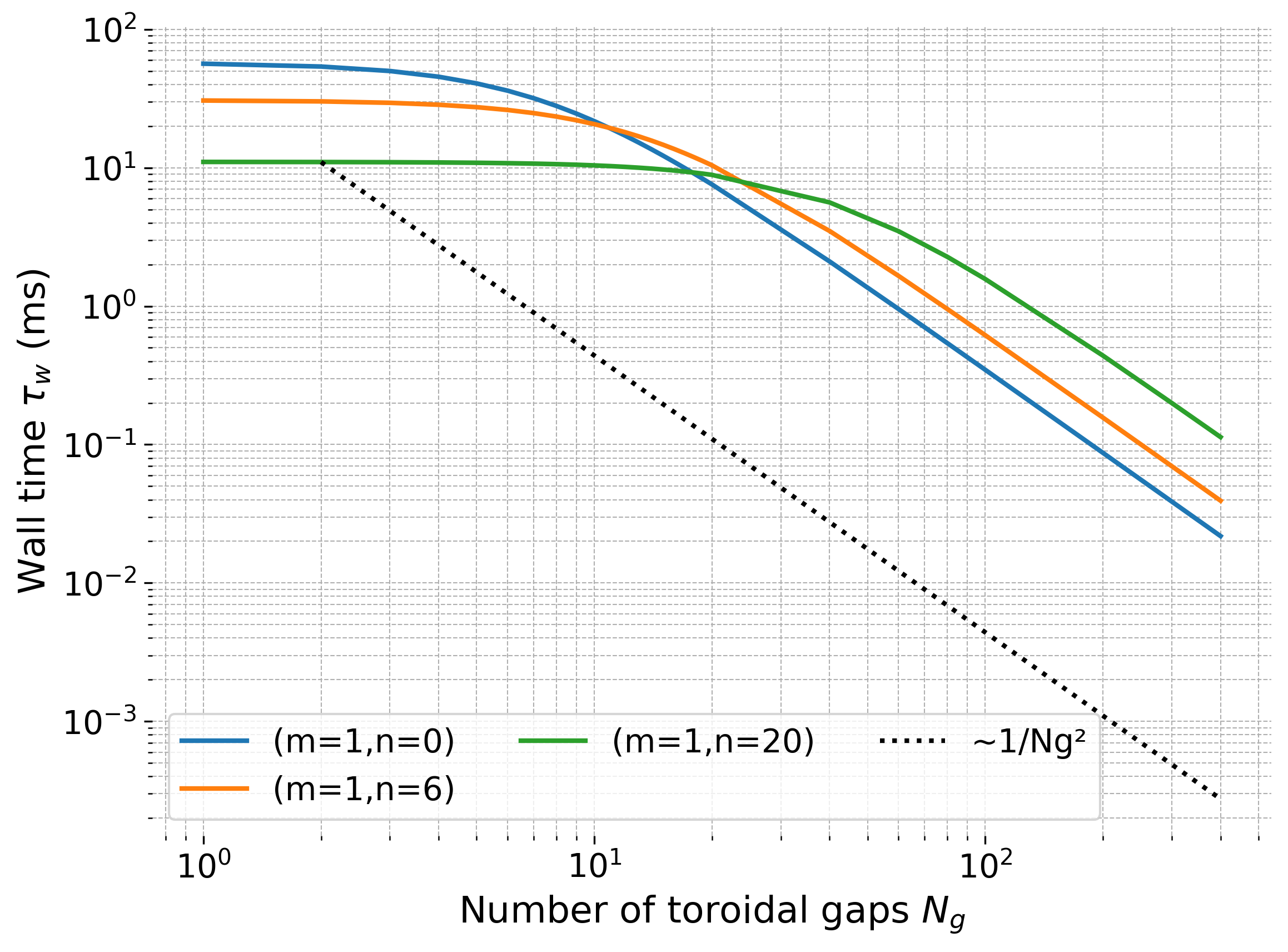}
    \caption{Scaling law for the wall time $\tau_w$ as a function of the number of gaps $N_g$ for different $m=1,n$ harmonics.}
    \label{fig:walltime_scaling}
\end{figure}

In particular, the relation in \eqref{eq:walltime_arma} allows quantifying the impact of
segmentation (Table~\ref{tab:results_compact} and Fig.~\ref{fig:walltime_ok}) by examining the competition between the two toroidal terms in the
denominator, namely $n^2$ and $N_g^2/4$. These represent the intrinsic toroidal
scale of the $n$ harmonic and the segmentation-imposed scale, respectively;
their relative magnitude determines when segmentation begins to influence the
wall time as shown for larger values of $N_g$ in Table \ref{tab:results_extended} and Fig. \ref{fig:walltime_scaling}.  Modes with low toroidal number ($n \ll N_g/2$) are affected early, and their wall time decreases rapidly with $N_g$. Intermediate-$n$ modes ($n \sim N_g/2$) retain the unsegmented behaviour until the segmentation scale
becomes comparable to the intrinsic one, after which the expected $1/N_g^2$
scaling is recovered. High-$n$ modes ($n \gg N_g/2$) remain dominated by their
intrinsic toroidal scale over a wide range of $N_g$, and segmentation becomes
relevant only when $N_g$ is increased substantially. This follows
directly from \eqref{eq:eta_eff_armadillo} and allows to assess when segmentation begins to influence a given $(m,n)$ harmonic.

\begin{table}[tb!]
    \centering
    \caption{Wall time $\tau_w$ (ms) for $m=1$ and representative toroidal mode numbers $n=0,6,20$ as a function of the number of toroidal gaps $N_g$.}
    \begin{tabular}{c|ccccccccccc}
	$n$ & $N_g=0$ & $2$ & $10$ & $20$ & $40$ & $60$ & $80$ 
	& $100$ & $200$ & $300$ & $400$ \\
	\hline
	0  & 57.39 & 53.85 & 21.71 & 7.58 & 2.10 & 0.95 & 0.54 
	& 0.35 & 0.087 & 0.039 & 0.022 \\
	6  & 30.85 & 30.26 & 20.73 & 10.45 & 3.50 & 1.66 & 0.96 
	& 0.62 & 0.157 & 0.070 & 0.039 \\
	20 & 10.85 & 10.82 & 10.23 & 8.74 & 5.53 & 3.43 & 2.24 
	& 1.55 & 0.433 & 0.197 & 0.111 \\
\end{tabular}
    \label{tab:results_extended}
\end{table}

While $m=1$ modes are the most relevant for RFP configurations, higher-$m$
harmonics become important in tokamaks, where $m\geq 2$ components are present in error-field spectra and RWM dynamics. The behaviour for higher
poloidal mode numbers follows directly from the effective
resistivity in \eqref{eq:eta_eff_armadillo}. In the denominator of
\eqref{eq:walltime_arma}, the intrinsic poloidal scale enters through the term
$m^2$, while the segmentation contribution depends only on the toroidal
combination $n^2 + N_g^2/4$. Increasing $m$ therefore enhances the intrinsic
poloidal scale without modifying the segmentation-imposed scale, reducing the
relative impact of segmentation for $m>1$. For example, for $m=2$ (Fig.~\ref{fig:walltime_ok} right) the intrinsic
contribution is four times larger than for $m=1$ at low $n$, suppressing the influence of
the $N_g^2$ term by approximately the same factor. As a consequence, a larger
number of gaps is required for the segmentation contribution to compete with
the intrinsic scale. The precise threshold depends on the combined intrinsic
structure $m^2+n^2$: for modes with small $n$ the required $N_g$ grows roughly
with $m$, whereas for modes with sufficiently large $n$ the toroidal term $n^2$ dominates and the segmentation must compete with it. If this condition is not met, the
reduction of the wall time with $N_g$ is correspondingly weaker than in the
$m=1$ case.

The analytical wall-time expression can also be used as a practical bridge between
normalized MHD stability analyses - where the dispersion relation naturally involves the combination $\gamma\tau_{\mathrm{w}}$ -  and growth rates expressed in physical units by using the mode-segmentation-dependent $\tau_w$ given by  \eqref{eq:walltime_arma}. This is essential for the design and implementation of MHD control schemes based on realistic time scales.

\begin{figure}[tb!]
    \centering
    \includegraphics[width=.8\textwidth]{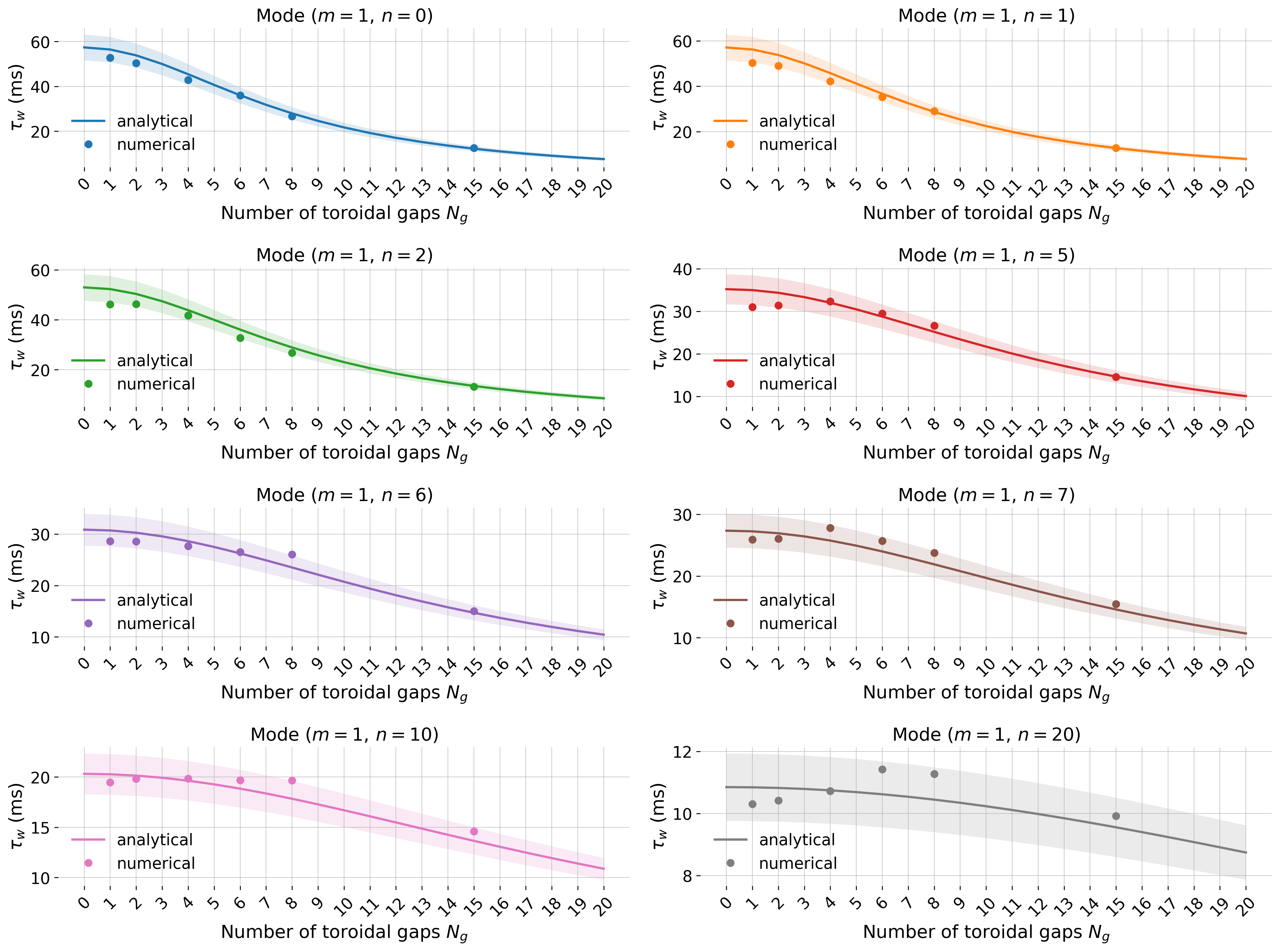}
\caption{Comparison of the analytical model results with the CARIDDI calculations for different $(m=1,n)$ and number of gaps $N_g$. The shaded area represents a $\pm 10\%$ variation around the analytical prediction.}
    \label{fig:walltime_ok_cariddi}
\end{figure}

For each $(m,n)$ and $N_g$, the analytical predictions are compared with the
numerical values obtained using the 3D electromagnetic model given by CARIDDI code~\cite{albanese1997finite}. In particular, the numerical wall time of a given harmonic is extracted from the eigenvalue analysis of the conducting structures. The wall–current eigenmodes are computed, their Fourier content is evaluated, and the modes carrying the largest amplitude of the desired $(m,n)$ harmonic are identified. Since no eigenmode of a fully 3D structure is a pure $(m,n)$ component, several eigenmodes contribute to the same harmonic with different weights. The corresponding decay rates (the eigenvalues) are therefore combined through a weighted average, with weights proportional to the $(m,n)$ content of each eigenmode, to obtain the numerical estimate of $\tau_w$. The comparison between analytical and numerical results are shown in Fig.~\ref{fig:walltime_ok_cariddi}: the accordance is within approximately $10\%$ (shaded area) across the considered range of $n$ and $N_g$ confirming that the segmentation effect is accurately described by the additional quadratic term in  \eqref{eq:walltime_arma}.

\section{Conclusions}
\label{sec:conclusions}

An analytical expression for the diffusion wall time of a toroidally segmented
conducting shell has been derived, extending the continuous–shell formulation to
account for the presence of $N_g$ toroidal gaps. Segmentation introduces an
additional toroidal scale that competes with the intrinsic structure of each
$(m,n)$ harmonic, leading to a mode‑dependent reduction of the wall time. The analysis reveals a clear hierarchy in the sensitivity of different
harmonics to segmentation. Modes with weak intrinsic toroidal structure
($n \ll N_g/2$) are affected early and exhibit a rapid decrease of the wall time
as the number of gaps increases. Modes with intermediate toroidal content
($n \sim N_g/2$) retain the behaviour of a continuous shell until the
segmentation scale becomes comparable to their intrinsic one, after which the
expected inverse‑quadratic scaling is recovered. High‑$n$ harmonics
($n \gg N_g/2$) remain dominated by their intrinsic toroidal scale over a wide
range of gap numbers, and segmentation becomes relevant only for sufficiently
large $N_g$. A similar trend applies to higher poloidal harmonics: increasing
$m$ strengthens the intrinsic poloidal scale and correspondingly reduces the
relative impact of segmentation. The analytical predictions have also been benchmarked against full 3D electromagnetic
calculations, showing agreement within approximately $10\%$ across the explored
range of mode numbers and gap configurations.

The resulting analytical formula provides a compact and physically transparent
tool for estimating the wall time of segmented conducting structures surrounding
the plasma. It can support design choices such as the number of gaps, shell
thickness, or material selection, and offers a fast alternative to full 3D
numerical simulations. The analytical formula allows also a direct conversion of RWM stability analysis results into growth rates in physical units, which are useful for designing feedback control systems and assessing stability margins. Although developed for an RFP configuration, the formulation is general and applicable to any confinement system where non‑axisymmetric conducting structures influence MHD stability and control, such as error‑field correction and RWM dynamics in tokamaks. 

\section*{Acknowledgements}
The authors thank L. Marrelli and R. Cavazzana for conceiving the original Armadillo concept, and R. Paccagnella and P. Zanca for their work on diffusion wall times for fields of general helicity. They also acknowledge G. Marchiori for contributions on the harmonic decomposition of state‑space eigenvectors. A particular thanks goes to V. D. Pustovitov for reading the entire manuscript and for his comments, which improved the completeness and quality of the paper, as well as for his continuous encouragement and support.

\section{References}
\bibliography{refs}
\bibliographystyle{unsrt}

\end{document}